\documentstyle[preprint,aps]{revtex}

\begin{document}
\draft
\preprint{26 July 1997}
\title{Low-Lying Excitations in the $S=1$ Antiferromagnetic Heisenberg  
       Chain}

\author{Shoji Yamamoto$^{\rm a,}$\footnote{E-mail: 
        yamamoto@hakuba.phys.okayama-u.ac.jp.},
        Seiji Miyashita$^{\rm b,}$\footnote{E-mail:  
miya@ess.sci.osaka-u.ac.jp.}}

\address{$^{\rm a}$Department of Physics, Faculty of Science,
         Okayama University, Tsushima, Okayama 700, Japan\\
         $^{\rm b}$Department of Earth and Space Science,
         Graduate School of Science, Osaka University,
         Toyonaka, Osaka 560, Japan}

\date{Received \hspace{5cm}}

\maketitle
\begin{abstract}
In order to confirm the picture of domain-wall excitations in the hidden
antiferromagnetic order of the Haldane phase, the structure of
the low-lying excitations in the $S=1$ antiferromagnetic Heisenberg chain
is studied by a quantum Monte Carlo method.
It is confirmed that there exists a finite energy gap between the first-
and the second-excited states at $k=\pi$ as well as between the ground
state and the first-excited state at $k=\pi$.
In the thermodynamic limit, the second-excited state at $k=\pi$ is
separated from the ground state by the gap which is three times as large
as the Haldane gap.
From the size dependences of the low-lying-excitation energies, the
interactions between 
the elementary excitations in the excited states are concluded to be
repulsive.
\end{abstract}


\narrowtext

   The structure of low-lying excitations of the $S=1$
antiferromagnetic Heisenberg chain is one of the main interests in the
study on the Haldane system.~\cite{Hald1,Hald2}
The valence-bond-solid (VBS) model~\cite{Affl1,Affl2} introduced  
by Affleck, Kennedy, Lieb, and Tasaki (AKLT), which gave a clear-cut  
physical picture of the Haldane massive state, stimulated several 
authors to study the nature of the excited states.
In order to describe the elementary excitations, Knabe~\cite{Knab1}
considered a triplet bond constructed from two spin-$1/2$ degrees of  
freedom in the VBS background, which is now called a crackion, 
while Arovas, Auerbach, and Haldane~\cite{Arov1} discussed a domain 
wall in the  hidden antiferromagnetic order.~\cite{Nijs1}
Later, F\'ath and S\'olyom~\cite{Fath1} demonstrated that both of the  
defects have a solitonic nature and actually give the same dispersion  
relation that well reproduces the elementary excitation spectrum of 
the AKLT model.~\cite{Fath1,Scha1}
On the other hand, proposing a new Monte Carlo technique,
Takahashi~\cite{Taka1} pioneeringly calculated the lower edge of the
excitation spectrum as a function of momentum $k$ for the Heisenberg  
chain.
There he further suggested that the lowest excitation of low momentum
may be a scattering state of two elementary excitations of momentum
$k\sim\pi$.
Motivated by his calculation and based on an idea of the hidden domain
wall, G\'omez-Santos~\cite{Gome1} made a variational approach to the  
Heisenberg chain and showed that the lowest excitations are
single-particle-like in the vicinity of $k=\pi$, whereas
two-particle-like near $k=0$.
Thus the lower edge of the excitation spectrum has almost fully been
investigated so far and it is probable that the elementary excitations
are more or less identified with the moving hidden domain wall.

   Although nowadays the higher excitation with arbitrary momentum is
generally believed to be a scattering state of the elementary domain  
walls,
quantitative investigation of them has not yet been performed well.
Developing quite different numerical treatments, White and
Huse~\cite{Whit1} and the present authors~\cite{Yama1} qualitatively
pointed out that the lowest excitations as a function of $k$ constitute
an isolated band at large $k$'s, while they coincide with the lower edge
of the two-domain-wall continuum at small $k$'s (Fig.~\ref{F:Illust}).
Their suggestion motivates us to calculate the second-lowest  
excitations.
Making use of the Lancz\"os method, Takahashi~\cite{Taka2} calculated
$S(q,\omega)$ of the Heisenberg chain of short length as a series of
$\delta$-function peaks at each excited state, which suggests that the
lowest excitations are actually separated from the upper continuum at large
$q$'s.  
For the AKLT model he found that the excitation energy of the
second-lowest state at $k=\pi$ is three times as large as one of the
lowest state at $k=\pi$.
However, the chain length $L \le 20$ he treated was not long enough to
confirm the same scenario in the thermodynamic limit at the Heisenberg
point where the correlation length is much longer than that of the AKLT
model.
Thus the expected multi-domain-wall excitation energies in the
thermodynamic limit, 
\begin{equation}
   E_2(0)-E_{\rm G}=2\Delta\,,\ \ 
   E_2(\pi)-E_{\rm G}=3\Delta\,,
   \label{E:gap}
\end{equation}
were not definitely obtained.
Here, $E_l(k)$ is the $l$~th eigenvalue in the $k$-momentum space,
$E_{\rm G}\equiv E_1(0)$ is the ground-state energy, and
$\Delta\equiv E_1(\pi)-E_{\rm G}$ is the Haldane gap.

   In order to study the structure of the low-lying energy levels, we
exploited an efficient quantum Monte Carlo method.
In principle the structure factor $S(q,\omega)$ can be obtained 
from the spin correlation function along the Trotter axis, $C(\tau)$,
which is called the imaginary time correlation function.
However, due to an intrinsic difficulty in the numerical treatment of
transforming the $C(\tau)$ to $S(q,\omega)$, certain statistical methods
such as the maximum-entropy technique,~\cite{Mesh1,Deis1,Silv1} 
have to be introduced for practical calculations at finite temperatures.
On the other hand, the lower boundary of the spectrum is simply
obtained from a logarithmic plot (log plot) of $C(\tau)$.
For the Haldane systems this simple approach works well and a precise
value of the lowest state at each momentum has been
obtained.~\cite{Yama2,Yama3,Yama4,Yama5}
Let us outline the log-plot method.

   We consider a periodic chain of $L$ spins described by the  
Hamiltonian
\begin{equation}
   {\cal H}=J\sum_{j=1}^L
      \mbox{\boldmath$S$}_{j} \cdot \mbox{\boldmath$S$}_{j+1}\,;\ \
      \mbox{\boldmath$S$}_{L+1}=\mbox{\boldmath$S$}_{1}\,,
   \label{E:H}
\end{equation}
where $\mbox{\boldmath$S$}_{j}$ is the $S=1$ spin operator.
Let us denote the $l$~th eigenvector and eigenvalue of ${\cal H}$ in  
the $k$-momentum space by
$\vert l;k\rangle$ and $E_{l}(k)$ [$E_1(k) \leq E_2(k) \leq \cdots$],
respectively.
When the Hamiltonian has the translational symmetry, i.e., $[{\cal H},
{\cal T}]=0$,
\begin{equation}
   {\cal H}\,\vert l;k\rangle=E_{l}(k)\,          \vert l;k\rangle  
\,,\ \
   {\cal T}\,\vert l;k\rangle={\rm e}^{{\rm i}k}\,\vert l;k\rangle  
\,,
   \label{E:eigen}
\end{equation}
where ${\cal T}$ is the translation operator.
The dynamic structure
factor in a real frequency domain, $S(q,\omega)$, is the 
Laplace transformation of the imaginary-time spin correlation
function of 
$S_q^z=L^{-1}\sum_{j=1}^L S_j^z{\rm e}^{{\rm i}qj}$:
\begin{equation}
   S(q,\tau)=\left\langle
   {\rm e}^{{\cal H}\tau}S_q^z{\rm e}^{-{\cal H}\tau}S_{-q}^z
             \right\rangle,
\end{equation}
where $\left\langle A\right\rangle \equiv
 {\rm Tr}[{\rm e}^{-\beta{\cal H}} A]/{\rm Tr}[{\rm e}^{-\beta{\cal  
H}}]$ denotes the canonical average at a given temperature  
$\beta^{-1}=k_{\rm B}T$.
When the system has some conserved quantities, the Hamiltonian is
block-diagonalized.
Since the total magnetization, $M=\sum_i S^z_i$, is a conserved quantity
in the present system, $S(q,\tau)$ is independently defined in each
subspace with a given $M$.
Using the complete vector set $\vert l;k\rangle$ in each subspace, 
$S(q,\tau)$ is  represented as
\begin{equation}
   S(q,\tau)
     =\frac{\sum_{l,l',k}{\rm e}^{-\beta E_{l\,}(k)}
      \left\vert \langle l;k \vert
      S_q^z
      \vert l';k+q \rangle \right\vert^2
      {\rm e}^{-\tau\left[E_{l'}(k+q)-E_{l}(k)\right]}}
      {\sum_{l,k}{\rm e}^{-\beta E_{l}(k)}}\,.
   \label{E:Sqt2}
\end{equation}
Thus $S(q,\tau)$ as a function of $\tau$ generally exhibits a  complicated
multi-exponential decay.
At a sufficiently low temperature, $S(q,\tau)$ is given as  
\begin{equation}
   S(q,\tau)=\sum_{l}
             \left\vert \langle 1;k_0 \vert
             S_q^z
             \vert l;k_0+q \rangle \right\vert^2
             {\rm e}^{-\tau\left[E_{l}(k_0+q)-E_1(k_0)\right]} \,,
   \label{E:Sqt3}
\end{equation}
where $k_0$ is the momentum at which the lowest state in the subspace  
is
located.
Now it is reasonable to approximate $E_1(k_0+q)-E_1(k_0)$ by the  
slope
$-\partial{\rm ln}[S(q,\tau)]/\partial\tau$ in the large-$\tau$  
region
satisfying
\begin{equation}
   \tau[E_2(k_0+q)-E_1(k_0+q)]\gg
   {\rm ln}
   \frac{\vert\langle 1;k_0 \vert S_q^z\vert n;k_0+q\rangle\vert^2}
        {\vert\langle 1;k_0 \vert S_q^z\vert 1;k_0+q\rangle\vert^2}\,,
   \label{E:cond}
\end{equation}
for an arbitrary $n$.
When the excitations constitute a isolated band and its spectral weight 
$\vert\langle 1;k_0 \vert S_q^z\vert 1;k_0+q\rangle\vert^2$ is large,
the inequality (\ref{E:cond}) is well justified in a wide region of $\tau$.
In fact these conditions are satisfied in the Haldane systems.
That is why the method did work quite well for the lowest excitations  
of the present system especially in the large-$q$ region.~\cite{Yama2}

   Taking a large enough Trotter number, we might, in principle, extract
the higher-lying levels from the log plots of $C(\tau)$ for proper regions
of $\tau$.
However, such an attempt has turned out to be unfeasible with the present
numerical facility.
Thus we use the log-plot method for the lowest level at each $q$,
but in the subspaces with various values of $M$.
Combining the calculations in these subspaces, we can construct the
low-energy structure and obtain the higher-lying levels.
Figure~\ref{F:Illust} illustrates the probable spectrum of the  
low-lying states of the system, which is based on the qualitative
arguments~\cite{Whit1,Yama1} and the calculations for the short  
chains of $L\leq 20$.~\cite{Taka2}

   In Fig.~\ref{F:Sqt} we show log plots of $S(q,\tau)$ calculated
in the subspaces of $M=0$, $M=1$, and $M=2$ as a function of $\tau$ at
various values of $q$, where the lattice constant was set equal to unity.
It seems that in general ${\rm ln}[S(q,\tau)]$ shows a better  
linearity at large $q$'s than at small $q$'s.
We have performed at least a million Monte Carlo steps to obtain
$S(q,\tau)$. In most cases 
the temperature $(\beta J)^{-1}$ and the Trotter number $n$ have been
set equal to $0.02$ and $200$, respectively.
We have checked that the thus-obtained data are reliable enough to
represent the ground state properties.
The numerical precision of the raw data amounts to two digits or more.

   As has been mentioned in our previous work,~\cite{Yama1}
a single spin flip create a domain wall of the hidden order parameter.
Thus the lowest-excitation energy is the gap between the lowest singlet
state ($S=0$) and the lowest triplet state ($S=1$).
Because the $S=0$ state contains a level of $M=0$, the gap immediately
above the ground state is obtained as the energy difference between the
two lowest states in the $M=0$ subspace except for the case of $q=0$.
For $q=0$, $S_q^z$ commutes with the Hamiltonian and therefore $S(q,\tau)$
does not depend on $\tau$.
The excitation energies from the ground state for the $M=0$ states are
obtained as
\begin{equation}
   \Delta E_0(q)=\Delta_0(q)\ \ (q\neq 0)\,.
   \label{delta00}
\end{equation}
Here $\Delta E_m(q)$ is the energy difference between the lowest-excited
state with a momentum $q$ in the $M=m$ subspace and the ground state, and
$\Delta_m(q)$ is the present numerical finding,
$-\partial{\rm ln}[S(q,\tau)]/\partial\tau$,
in the subspace of $M=m$ under the condition (\ref{E:cond}).
Takahashi showed in his recent work~\cite{Taka2} that the spectral
weight of the lowest excitation at each $q$ is extremely large especially 
at $q\agt 0.3\pi$ where the lowest excitation is separated from the upper
continuum.~\cite{Gome1,Taka2}
This fact causes the fine linearity of ${\rm ln}[S(q,\tau)]$ at large
$q$'s but the less fine linearity at small $q$'s, as was observed in
Fig.~\ref{F:Sqt}(a).
In Fact, in our previous attempt~\cite{Yama2} to obtain the lower edge
of the spectrum, the data at small $q$'s were not so satisfying as ones
at large $q$'s from the point of the numerical precision.
We note that the inelastic-neutron-scattering measurements~\cite{Ma1} 
actually revealed the clear isolated band for the $S=1$ Haldane material
Ni(C$_2$H$_8$N$_2$)$_2$NO$_2$(ClO$_4$).

   In the subspace of $M=1$, the present approach brings us the energy
differences between the lowest triplet state ($k=\pi$) and other triplet
states with an arbitrary momentum.
The excitation energies from the ground state for the $M=1$ states are
obtained as
\begin{equation}
   \Delta E_1(q)=\Delta_0(\pi)+\Delta_1(\pi-q)\,.
   \label{delta10}
\end{equation}
In principle, $\Delta E_0(q)$ obtained from the calculation for $M=0$
should agree with $\Delta E_1(q)$ obtained from one for $M=1$.
However, a certain difficulty in the present treatment prevents us from
reaching the definite coincidence between them.
In the $M=0$ subspace, the lowest state at each $q$ has a relatively large
spectral weight, while in the $M=1$ subspace, this seems not to be the
case.
Actually, for $M=1$, the effect of the multi-exponential decay more
clearly appears in the log plot of $S(q,\tau)$ especially at small $q$'s.
But, even in the case of $M=1$, the log plot of $S(q,\tau)$ still gives
an almost straight line in the vicinity of $q=\pi$, which results in a
precise estimate of
$\Delta E_0(0)=\Delta E_1(0)=\Delta_0(\pi)+\Delta_1(\pi)$.

   In the subspace of $M=2$, the present calculation brings us
the energy differences between the lowest quintuplet state
($k=0$)~\cite{Fath1} and other quintuplet states with an arbitrary
momentum.
The excitation energies from the ground state for the $M=2$ states are
obtained as
\begin{equation}
   \Delta E_2(q)=\Delta E_0(0)+\Delta_2(q)
                =\Delta_0(\pi)+\Delta_1(\pi)+\Delta_2(q)\,.
   \label{delta20}
\end{equation}
Here nonlinearity of ${\rm ln}[S(q,\tau)]$ persists in a relatively wide
range of $\tau$ even at large $q$'s, which prevents us obtaining a full
dispersion curve of the second-excited states.

   We plot in Fig.~\ref{F:dsp} the lowest and the second-lowest
eigenvalues, $\Delta E_1(q)$ and $\Delta E_2(q)$, as a function of $q$
for the $L=64$ chain, which have been obtained using $\Delta_0(\pi)$ and
$\Delta_1(\pi)$ at $L=64$.
Here all the errors arise in estimating the slope of
${\rm ln}[S(q,\tau)]$ rather than come from the raw Monte Carlo data.
The lowest eigenvalues $\Delta E_1(k)$ were determined, using both of
$S(q,\tau)$'s with $M=0$ and $M=1$, so as to minimize the numerical
ambiguity.
Although $\Delta E_2(q)$ has not successfully been obtained in the
region of $q\alt 0.7\pi$, we here clearly confirm the existence of the
isolated band and observe the lower edge of the three-domain-wall
continuum as well as one of the two-domain-wall continuum.
The overall behavior is almost the same as that for $L=128$ except
in the vicinity of $q=0$ and $q=\pi$, and therefore we can believe that we
are observing the bulk behavior.

   In Fig.~\ref{F:gap} we show size dependences of the first-excitation
energy at the zone boundary, $\Delta E_0(\pi)=\Delta E_1(\pi)$,
the first-excitation energy at the zone center,
$\Delta E_1(0)=\Delta E_2(0)$, and the second-excitation energy,
$\Delta E_2(\pi)$, which are supposed to be the bottom of the
single-domain-wall band, one of the two-domain-wall continuum,
and one of the three-domain-wall continuum, respectively.
Here the symbols $\times$ represent the Takahashi's  
data~\cite{Taka2} for $\Delta E_2(\pi)$ obtained through a different 
method, which are somewhat inconsistent with our finding 
beyond the numerical uncertainty.
We here observe that the relation (\ref{E:gap}) comes to hold as $L$
increases.
Therefore, we conclude that the low-lying excitations of the present
model are regarded as the domain-wall excitations in the hidden
antiferromagnetic order.

   The chain-length dependences of the energy of the
domain-wall-scattering state suggest that there exists a long-range
repulsive interaction between the domain walls in the excited states,
which is qualitatively consistent with a variational
calculation.~\cite{Neug1}
We note that the interaction between the domain walls is contrastingly
attractive in the ground state.~\cite{Yama6}
In the AKLT model,~\cite{Taka2} the relation (\ref{E:gap}) almost holds
even at $L=16$.
It is well known that the spin-spin correlation length is
$1/{\rm ln}3\simeq 0.91$ at the AKLT point,~\cite{Affl2} while it is
estimated to be $6.2$ at the Heisenberg point.~\cite{Goli1}
Thus the spatial extension of the domain wall is expected to be much
larger in the Heisenberg model than in the AKLT model, to which the
present significant size dependence is attributed. 
Smearing the domain wall over three lattice sites, Scharf and
Mikeska~\cite{Scha1} obtained a variational bound for the Haldane gap
of the AKLT model which coincides with the exact-diagonalization result
within 1\% error.
The spatial extension of the domain wall in the Heisenberg model may
reach more than twenty lattice sites.

\acknowledgments

   The authors wish to thank Professor M. Kaburagi for his suggestion for
the present study.
They also thank Professor H.-J. Mikeska for stimulating discussion with
him.
Numerical calculations were mainly carried out using the facilities  
of the Supercomputer Center, Institute for Solid State Physics,  
University of Tokyo.
The present work is partly supported by Grants-in-Aid of 
the Ministry of Education, Science, and Culture, and by a Grant-in-Aid
of the Okayama Foundation for Science and Technology.

\begin{figure}
   \caption{Illustration of the spectrum of the low-lying excited states
            as a function of momentum for the infinite chain.}
   \label{F:Illust}
\end{figure}

\begin{figure}
   \caption{Logarithmic plots of $S(q,\tau)$ versus $\tau$ at various  
            values of $q$ for the $L=64$ chain: (a) $M=0$, (b) $M=1$,
            and (c) $M=2$.}
   \label{F:Sqt}
\end{figure}

\begin{figure}
   \caption{The lowest and the second lowest eigenvalues as a function
            of $q$ for the $L=64$ chain, where $\bigcirc$, $\Diamond$,
            and $\Box$ denote the results obtained from $S(q,\tau)$
            calculated under $M=0$, $1$, and $2$, respectively.}
   \label{F:dsp}
\end{figure}

\begin{figure}
   \caption{Size dependences of the excitation energy of the
            first-excited state at the zone boundary, $\Delta E_0(\pi)$,
            one of the first-excited state at the zone center,
            $\Delta E_1(0)$, and one of the second-excited state at the
            zone boundary, $\Delta E_2(\pi)$.
            The black symbols represent the $L\rightarrow\infty$
            extrapolated values.
            The symbols $\times$ represent the Takahashi's calculations
            [14] for $\Delta E_2(\pi)$ obtained with an
            exact-diagonalization technique.}
   \label{F:gap}
\end{figure}
\end{document}